**_N_-type Conversion of SnS by Isovalent Ion Substitution: Geometrical Doping As a New Doping Route**

*Fan-Yong Ran, Zewen Xiao, Yoshitake Toda, Hidenori Hiramatsu, Hideo Hosono, and Toshio Kamiya**

Dr. F. -Y. Ran, Z. Xiao, Prof. H. Hiramatsu, Prof. H. Hosono, and Prof.T. Kamiya
Materials and Structures Laboratory,
Tokyo Institute of Technology
4259 Nagatsuta, Midori-ku, Yokohama 226-8503, Japan
E-mail: kamiya.t.aa@m.titech.ac.jp

Prof. H. Hosono
Frontier Research Center
Tokyo Institute of Technology
4259 Nagatsuta, Midori-ku, Yokohama 226-8503, Japan

Dr. Y. Toda, Prof. H. Hiramatsu, Prof. H. Hosono, and Prof.T. Kamiya
Materials Research Center for Element Strategy
Tokyo Institute of Technology
4259 Nagatsuta, Midori-ku, Yokohama 226-8503, Japan
E-mail: kamiya.t.aa@m.titech.ac.jp

Keywords: isovalent ion substitution, carrier polarity conversion, tin monosulfide (SnS), Pb subsitution

Tin monosulfide (SnS) is a naturally *p*-type semiconductor with a layered crystal structure, but no reliable *n*-type SnS has been obtained by conventional aliovalent ion substitution. In this work, we succeeded in carrier polarity conversion to *n*-type by isovalent ion substitution for polycrystalline SnS thin films on glass substrates. Substituting $Pb^{2+}$ for $Sn^{2+}$ converted the majority carrier from hole to electron, and the free electron density ranged from $10^{12}$ to $10^{15}$ $cm^{-3}$ with the largest electron mobility of 7.0 $cm^2/(Vs)$. The *n*-type conduction was confirmed further by electrical characteristics of pn heterojunctions. Density functional theory calculations revealed that the Pb substitution invokes a geometrical size effect that enlarges the interlayer distance and subsequently reduces the formation energies of Sn and Pb interstitials, which work as electron donors.

**1. Introduction**

Control of carrier polarity conversion in semiconductor is important to produce high-performance semiconductor devices such as solar cells and light emitters, and is actually utilized in conventional semiconductors such as Si and compound semiconductors. On the other hand, it is known that such bipolar doping is not attained easily in other semiconductors; e.g., most of oxide semiconductors are of naturally *n*-type, and it is difficult to obtain *p*-type conduction in the same materials as known e.g. for $SnO_2$, and vice versa e.g. for $Cu_2O$. To date, several, but a limited number of doping routes have been recognized and employed. For ionic semiconductors, aliovalent ion substitution and off-chemical stoichiometry are known well; e.g., substitution of $Zn^{2+}$ with $Ga^{3+}$ increased the electron density in ZnO,[1] and Cu vacancy increased the hole density in $Cu_2O$.[2] Further, H doping is now recognized as an important and effective route for *n*-type doping in oxide semiconductors.[3] For organic semiconductors and devices, chemical doping, which is caused by partial charge transfer originating from different electron affinities of constituent atoms / functional groups, is important.[4] The most popular route for carrier polarity conversion is aliovalent ion substitution; actually, *n*-type conversion of SnO was realized by substituting $Sb^{3+}$ ions for the $Sn^{2+}$ ions.[5] Up to now, however, this route has not been successful for many semiconductors, such as SnS.

SnS is a naturally *p*-type semiconductor with hole densities from $10^{15}$–$10^{18}$ $cm^{-3}$ and the high mobilities ~90 $cm^2/(Vs)$.[6,7] It has a layered crystal structure along the *a*-axis direction as shown in **Figure 1**a, which belongs to the orthorhombic lattice (the space group *Pnma*, No. 62). Due to its reasonable small bandgap of ~1.07 eV and strong optical absorption coefficients above the bandgap (> $10^5$ $cm^{-1}$), [8] SnS is expected to be a promising absorber material for low-cost thin-film solar cells. Thus, numerous *n*-type materials, including CdS,[9,10] $SnS_2$,[11] $FeS_2$,[12] $TiO_2$,[13] ZnO,[14] and a-Si,[15] have been employed for fabricating heterjunction SnS-based solar cells. However, the highest energy conversion efficiency reported up to now is limited to ~2.9%,[16,17] which is much lower than the theoretically-predicted value of 24%.[18] The low efficiency might suffer from unfavorable band alignments and the large lattice mismatches in the heterjunction structures.[19,20] Fabricating a homojunction solar cell with *p*-SnS/*n*-SnS structure would solve this problem.

With this line, much effort has been devoted to obtaining *n*-type SnS materials by substituting the $Sn^{2+}$ ions with aliovalent ions with the charge state of 3+. Dussan *et al*. report that $Bi^{3+}$-doped SnS exhibits *n*-type conduction when the Bi concentration is larger than 50%.[21] Whereas, a $Bi_2S_3$ impurity phase, which is also *n*-type, was observed in their heavily Bi-doped SnS films.[22] Sajeesheesh *et al*. claim that *n*-type SnS thin films are obtained by chemical spray pyrolysis, but their result might be due to a significant *n*-type $Sn_2S_3$ impurity phase in the films.[23] Very recently, Sinsermsuksakul *et al*. tried to obtain *n*-type SnS by $Sb^{3+}$ doping; however, except for great increase in the electrical resistance of the SnS film, no *n*-type conduction was observed.[24] That is, no reliable *n*-type SnS material has yet been reported.

In this work, we succeeded in fabricating reliable *n*-type SnS films by isovalent $Pb^{2+}$ doping. We found that the doping mechanism is strikingly different from the conventional doping routes such as ion substitution, off-stoichiometry, and chemical doping. Substitution for the $Sn^{2+}$ ion with a larger $Pb^{2+}$ ion increases the interlayer distance in SnS, and this geometrical effect induces the formation of Sn / Pb interstitials easier, and the interstitial ions work as donors.

## 2. Results and discussion

### 2.1. Crystal Structure

We deposited $(Sn_{1-x_f}Pb_{x_f})S$ films ($x_f$: film chemical composition) by pulsed laser deposition (PLD) on $SiO_2$ glass substrates in a $H_2S$ gas flow to control the chemical stoichiometry [i.e., the (Sn+Pb) : S ratio], where the $H_2S$ pressure ($P$) was a variable parameter. $(Sn_{1-x_t}Pb_{x_t})S$ polycrystalline disks with the target chemical composition $x_t$ = 0.18, 0.37, and 0.66 were used as ablation targets. X-ray fluorescence (XRF) spectroscopy confirmed that these $x_t$ produced thin films with $x_f$ = 0.08–0.5. The details of experimental and calculation can be found in supplementary information.

First, we confirmed the film structures by X-ray diffraction (XRD). **Figure 1**b shows a typical out-of-plane $2\theta$–$\omega$ synchronous scan (top panel) and an in-plane synchronous $2\theta_\chi$-$\phi$ scan (bottom panel) XRD patterns of the $(Sn_{1-x_f}Pb_{x_f})S$ film with $x_f$ = 0.5 grown at $T_s$ = 300 °C and $P$ = 5 Pa. The out-of-plane XRD pattern exhibited strong 200, 400 and 800 diffractions of the orthorhombic structure, which is the same as that of pure SnS in **Figure 1**a, along with a weak 011 diffraction. As seen in **Figure S1** (supplementary information), orthorhombic $(Sn_{1-x_f}Pb_{x_f})S$ films were obtained at $P \leq 15$ Pa (corresponding to the closed symbols in **Figure 1**c); while, amorphous films were obtained when $P$ was increased to 20 Pa (the open symbols in **Figure 1**c). The in-plane synchronous $2\theta_\chi$-$\phi$ scan (bottom panel) shows powder-like patterns with all the possible *hkl* diffractions, suggesting that the film did not have in-plane orientation. It was further confirmed by in-plane rocking patterns ($\phi$ scan at fixed $2\theta_\chi$, data not shown); all the data showed that the crystallized films did not have a preferential orientation in plane. These results indicate that the $(Sn_{1-x_f}Pb_{x_f})S$ films were polycrystalline films with a strong 100 preferential orientation normal to the substrate. No impurity phase was detected both in the out-of-plane and the in-plane XRD patterns.

**Figure 1**c shows the variation of $x_f$ as functions of $P$, $T_s$ and $x_t$. It is seen that all the $x_f$ values were smaller than the $x_t$ values of the corresponding targets, and the $x_f$ values decreased with increasing $P$. As seen for the target with $x_t$ = 0.37, the maximum amount of Pb was incorporated in the films at $T_s$ = 300 °C. We, therefore, employed $T_s$ = 300 °C hereafter.

The crystallized region in **Figure 1**c is classified further to three regions as indicated by the dashed lines. Region I is "*p*-type region" (high $P \geq 15$ Pa at low $x_f < 0.1$), where the films still exhibited *p*-type conduction with low hole densities ($N_h$) and low hole mobilities ($\mu_h$) (measured by Hall effect, details will be discussed for **Figure 2**). *n*-type ($Sn_{1-x_f}Pb_{x_f}$)S films were obtained in Region II ("*n*-type region", $x_f \geq 0.15$ at low $P \leq 10$ Pa). The electron density ($N_e$) and mobility ($\mu_e$) changed largely with $x_f$ and $P$, which will be discussed later on. Region III is the intermediate region ("highly-resistive region", low $x_f$ & low $P$, and high $x_f$ & high $P$), where the films exhibited very high resistivity >$10^5$ Ω·cm, and the Hall effect measurements did not give definite Hall voltage signs.

Here, we discuss the doping structure of Pb. **Figure 1**d shows the variation of the out-of-plane 400 diffraction angles $2\theta_{400}$ obtained by $2\theta-\omega$ synchronous scan as a function of $x_f$. The $2\theta_{400}$ value shifted to lower angles as $x_f$ increased, indicating that the *a*-axis expanded with increasing $x_f$. The lattice parameters obtained from the out-of-plane 400 and the in-plane 020 & 011 diffraction angles are summarized as a function of $x_f$ in **Figure 1**e. As $x_f$ increased from 0 to 0.5, the *a* and *b* values increased linearly from 1.122 to 1.148 nm and from 0.403 to 0.414 nm, respectively, whereas the *c* value decreased from 0.426 to 0.419 nm; i.e., the interlayer distance (corresponding to the *a* value) increased. The solid lines in **Figure 1**e represent the lattice parameters of the ($Sn_{1-x}Pb_x$)S bulk sample reported by Leute *et al.*.[25] The *a* values of our films are almost the same as those of the bulk samples. However, the *b* and *c* values exhibited non-negligible deviations from the bulk values; *i.e.*, the *b*-axis was expanded while the *c*-axis shrunken compared from the bulk values. The reason is not clear, but defects in the polycrystalline films would cause the structural difference. **Figure 1**e also compares the variation of the lattice parameters with those obtained by density functional theory (DFT) calculations (the open symbols) performed with the ($Sn_{16-n}Pb_n$)$S_{16}$ super cell model indicated by the black line box in **Figure 1**f. Here, local density approximation (LDA) and generalized gradient approximation (GGA) functionals are compared. As will be seen later, GGA provides better description about the electronic structure; however, here we can see that the experimental results were within the variation of the functionals. That is, this model, where the Sn sites are substituted by Pb, explains the experimental structure well, and strongly supports that the Pb dopants are successfully incorporated to the Sn sites in the SnS lattice.

**2.2. Electrical and Electronic Properties**

**Figure 2**a shows Hall effect measurement results as a function of $x_f$. The pure SnS film showed *p*-type conduction with $N_h$ ~ 4.1×10$^{15}$ cm$^{-3}$ and $\mu_h$ ~ 12 cm$^2$/(Vs). The $(Sn_{1-x_f}Pb_{x_f})S$ film with $x_f$ = 0.08 fabricated at $P$ = 15 Pa still showed *p*-type conduction but with the low $N_h$ ~ 1.0×10$^{14}$ cm$^{-3}$ and the very small $\mu_{Hall}$ in the order of 10$^{-2}$ cm$^2$/(Vs). When $x_f \geq 0.2$, *n*-type conductions were observed for the films fabricated at $P$ = 5 and 10 Pa. For the *n*-type $(Sn_{1-x_f}Pb_{x_f})S$ film with $x_f$ = 0.2, $N_e$ and $\mu_{Hall}$ were 1.4×10$^{12}$ cm$^{-3}$ and 1.3 cm$^2$/(Vs), respectively. $N_e$ increased with increasing $x_f$ and reached 2.0×10$^{15}$ cm$^{-3}$ for $x_f$ = 0.5. $\mu_e$ was not changed largely when $x_f$ < 0.3 ($N_e$ < 3.2×10$^{13}$ cm$^{-3}$). At $x_f$ values > 0.4, $\mu_e$ increased almost linearly, and the maximum value of 7.0 cm$^2$/(Vs) was obtained for $x_f$ = 0.5.

**Figure 2**b shows temperature dependences of $N_e$ and $\mu_e$ for the $(Sn_{1-x_f}Pb_{x_f})S$ film with the room-temperature $N_e$ of 4.3×10$^{13}$ cm$^{-3}$ ($x_f$ = 0.48 grown at 10 Pa). The $N_e$ shows a thermally-activated behavior with an active energy of $E_a$ ~ 0.4 eV. From a simple approximation in the impurity region $N_e = (N_D N_C)^{1/2}\exp[-(E_C - E_D)/(2k_BT)]$ ($N_D$ is the donor density, $N_C$ the conduction band effective density of states (DOS), $E_C - E_D$ the donor level measured from the conduction band minimum $E_C$, $k_B$ the Boltzmann constant), $E_C - E_D$ and $N_D$ are estimated to be ~0.8 eV and 2.8×10$^{22}$ cm$^{-3}$, respectively. More accurate estimation was performed based on the total DOS obtained by the DFT calculation and the semiconductor statistics,[26] which provided $E_C$ - $E_D$ = 0.75 eV, $E_C - E_F$ = 0.30 eV, and $N_D$ = 1.0×10$^{21}$ cm$^{-3}$, agreeing well with the above simple estimation and guaranteeing that the film is in the impurity region in this measurement temperature range. On the other hand, although $E_F$ was closer to $E_C$ as in usual *n*-type semiconductors, the donor level $E_D$ was closer to the valence band maximum energy

($E_V$) rather than $E_C$, showing that the *n*-type doping in the $(Sn_{1-x_f}Pb_{x_f})S$ films is a bit different from the usual *n*-type semiconductors.

As shown by the red line in **Figure 2**b, $\mu_e$ decreased with decreasing the temperature, and the $\ln(\mu_{Hall}T^{1/2}) - T^{-1}$ plot exhibited a good straight line in the whole $T$ range, suggesting that the electron transport in the film was dominated by grain boundary (GB) potential barriers as proposed by Seto *et al.*,[27] where electron transport is disturbed by potential barriers formed due to the electrons trapped at acceptor-type defects at the GBs. The GB potential barrier height $E_B$ is estimated to be approximately 0.09 eV (the equation is given in **Figure 2**b)[27]. From this result, we can estimate the potential electron mobility $\mu_0$ (i.e., the ideal value when no GB affects the carrier transport) by extrapolating $E_B$ to zero (*i.e.*, $\mu_0 = \mu_{Hall} \exp(E_B/kT)$), which gives $\mu_0 \sim 1.6\times10^2$ cm$^2$/(Vs).

**Figure 2**c shows a valence band structure of a $(Sn_{0.5}Pb_{0.5})S$ film measured by ultraviolet photoemission spectroscopy (UPS). A sharp peak at 1 – 2 eV and a broad peak at 2.5 – 4.5 eV can be observed, agreeing with the projected DOS (PDOS) calculated by DFT in **Figure 2**e. The valence band consists mainly of S 3*p* orbitals, which slightly hybridized with Sn 5*s*, Sn 5*p*, Sn 5*d*, Pb 6*s*, Pb 6*p*, and Pb 6*d* orbitals. As seen in **Figure 2**d, the observed $E_F$ of the $(Sn_{0.5}Pb_{0.5})S$ film is located at 0.82 eV above VBM. From the bandgap value of 1.15 eV (supplementary information **Figure S**2), the $E_C - E_F$ value is estimated to be 0.33 eV, closer to conduction band maximum (CBM).

To further confirm the *n*-type conduction of these films, *n*-type $(Sn_{0.5}Pb_{0.5})S$ / *p*-type Si pn heterjunction was prepared (the device structure is shown in the inset to **Figure 2**f). The *n*-$(Sn_{0.5}Pb_{0.5})S$ film and the *p*-Si wafer used had $N_e = 2\times10^{15}$ and $N_h = 5\times10^{15}$ cm$^{-3}$, respectively. The current–voltage (*I*–*V*) characteristic of the pn junction (**Figure 2**f) showed a clear rectifying characteristic, further supporting the *n*-type conduction of the $(Sn_{1-x_f}Pb_{x_f})S$ film. The band alignment of this pn heterojunction (supplementary information **Figure S**3) gives the built-in potential ($V_{bi}$) of 0.76 eV. This $V_{bi}$ roughly explains the experimental threshold voltage of the pn heterojunction ~0.67 V obtained by extrapolating the straight line region in **Figure 2**f.

**2.3. Discussion**

Here, we like to discuss the origin of the *n*-type doping in the $(Sn_{1-x_f}Pb_{x_f})S$ films. It is known that Pb ions favor to take +2 and +4 valence states, and the latter would explain the *n*-type doping if $Pb^{4+}$ substitutes the $Sn^{2+}$ site. However, the above DFT calculations for the $(Sn_{32-n}Pb_n)S_{32}$ supercell models indicated that the Pb substitutions at the Sn site (denoted $Pb_{Sn}$) generate no free charges because the Pb is ionized to $Pb^{2+}$. We also confirmed by X-ray photoemission spectroscopy (XPS) that the calibrated energy level of Pb $4f_{7/2}$ in the $(Sn_{1-x_f}Pb_{x_f})S$ with $x_f$ = 0.5 was 137.55 eV and close to that in a reference PbS (137.15 eV) (supplementary information **Figure S**4), which supports that the charge state of the Pb incorporated in the $(Sn_{1-x_f}Pb_{x_f})S$ films is +2. This result in turn indicates that the doping mechanism by this Pb substitution is not an aliovalent ion substitution, and the conventional substitution models do not explain the *n*-type doping in the $(Sn_{1-x_f}Pb_{x_f})S$ films.

Here, we discuss the microscopic mechanism of the *n*-type doping by the Pb substitution. Firstly, we should remind that the *n*-type conduction was obtained only when a film was grown under the S-poor condition (*i.e.*, at low $H_2S$ pressures). We calculated the formation enthalpies ($\Delta H_f$) of intrinsic defects in the pure SnS and the $(Sn_{0.5}Pb_{0.5})S$ models (the blue line box in **Figure 1**f) under the S-poor limit condition as a function of $E_F$ by DFT calculations as shown in **Figures 3**a and b, respectively. Vacancies ($V_S$, $V_{Sn}$, $V_{Pb}$), anti-site defects ($Sn_S$, $Pb_S$), interstitials ($Sn_i$, $Pb_i$) were examined (see **Figure 1**f for the models) with the defect charge states from 2+ to 2-. These calculations employed LDA functionals not GGA in order to compare with the previously-reported results for pure SnS by Vidal *et al.*.[28] The present result of the pure SnS model (**Figure 4**a) is almost the same as their results; *i.e.*, the most stable charge state of $V_S$ transits from 2+ to 0 at $E_F$ ~ 0.4 eV, corresponding to the charge transfer energy level of $\varepsilon_{2+/0}$. The most stable defect changed from $V_S^{2+}$ to $Sn_S^-$ & $V_{Sn}^{2-}$ at $E_F$ ~ 0.4 eV. As $V_S^{2+}$ acts as a doubly-ionized donor while $Sn_S^-$ and $V_{Sn}^{2-}$ are ionized acceptors, suggesting that SnS is intrinsically compensated *p*-type semiconductor. For quantitative analysis, the equilibrium $E_F$ ($E_{F,e}$) at 300 °C (i.e., we assume the defect structures at the growth temperature were frozen to room temperature) was calculated by considering all the $\Delta H_f$ values and the semiconductor statistics, giving $E_{F,e} - E_V = 0.41$ eV with $[V_S^{2+}] = 5.0\times10^{15}$, $[V_{Sn}^{2-}] = 3.9\times10^{15}$, and $[Sn_S^-] = 5.2\times10^{15}$ cm$^{-3}$ for the SnS model. This means that the free electrons were generated from $V_S^{2+}$ at $1.0\times10^{16}$ cm$^{-3}$ but compensated by larger amounts of holes generated from $V_{Sn}^{2-}$ and $Sn_S^-$ at $1.3\times10^{16}$ cm$^{-3}$, resulting in the *p*-type conduction. It should be noted that the $Sn_i$ has a very large $\Delta H_f$ and is not likely formed in pure SnS.

For the $(Sn_{0.5}Pb_{0.5})S$ model in **Figure 3**b, although the $\Delta H_f$ values of $V_S^{2+,0}$ remained unchanged, that of $Sn_i^{2+}$ was reduced and that of $V_{Sn}^{2-}$ increased significantly compared to those in the pure SnS, which is because the interlayer distance (corresponding to the *a*-axis length) and the *b*-axis lattice parameters were increased by the Pb substitution (as also observed experimentally in **Figure 1**e). Similar $\Delta H_f$ behaviors were found also for $Pb_i^{2+}$ and $V_{Pb}^{2-}$, respectively. That means, this geometrical alternation makes the generation of the donor $Sn_i^{2+}$ and $Pb_i^{2+}$ easier whereas suppresses the generation of the acceptor $V_{Sn}^{2-}$ and $V_{Pb}^{2-}$, suggesting *n*-type doping. The $E_{F,e}$ calculation gave $E_{F,e} - E_V = 0.71$ eV with $[V_{Sn}^{2-}] = 5.5\times10^{13}$, $[V_{Pb}^{2-}] = 4.6\times10^{14}$, $[Sn_i^{2+}] = 6.1\times10^{14}$, and $[Pb_i^{2+}] = 3.2\times10^{13}$ cm$^{-3}$. Note that the $V_S$ has the charge neutral state at this $E_{F,e}$, and does not contribute to charge doping. Consequently, the free holes were generated at $1.0\times10^{15}$ cm$^{-3}$, while the larger amounts of free electrons were generated at $1.3\times10^{15}$ cm$^{-3}$, resulting in *n*-type doping. Finally, we conclude that $Sn_i$ and $Pb_i$ are the most plausible origin of the *n*-type conduction in the $(Sn_{x_f}Pb_{1-x_f})S$ films.

## 3. Conclusion

In summary, *n*-type conduction in SnS was achieved by isovalent Pb substitution with the maximum electron mobility of 7 $cm^2/(Vs)$. DFT calculations proposed a new doping model where the Pb substitution at the Sn sites induces the formation of $Sn_i$ and/or $Pb_i$ and produces donors. To date, carrier polarity control in semiconductor is achieved mainly by aliovalent ion substitution, off chemical stoichiometry, chemical doping and so on. This work revealed that substitution by an isovalent ion can also induce carrier doping by a two-step indirect mechanism through a geometrical effect and subsequent formation of charged defects. This way of thinking would provide more flexibility to explore new doping routes, open a new way for controlling carrier polarity and density in novel semiconductors in which conventional aliovalent ion substitution is difficult.

## Supporting Information

Experimental details, as well as additional XRD data, optical data, XPS data, and band alignment diagram of *n*-type $(Sn_{0.5}Pb_{0.5})S$ / *p*-type Si heterjunction can be found in supplementary information file which is available is available from the Wiley Online Library or from the author T.K.

## Acknowledgements

F.-Y. Ran and T. Kamiya were supported by Funding Program for Next Generation World-Leading Researchers (NEXT Program, Project #GR035). This work was supported also by the New Energy and Industrial Technology Development Organization (NEDO) under the Ministry of Economy, Trade and Industry (METI) and by the Element Strategy Initiative to Form Core Research Center and the Ministry of Education, Culture, Sports, Science and Technology (MEXT).

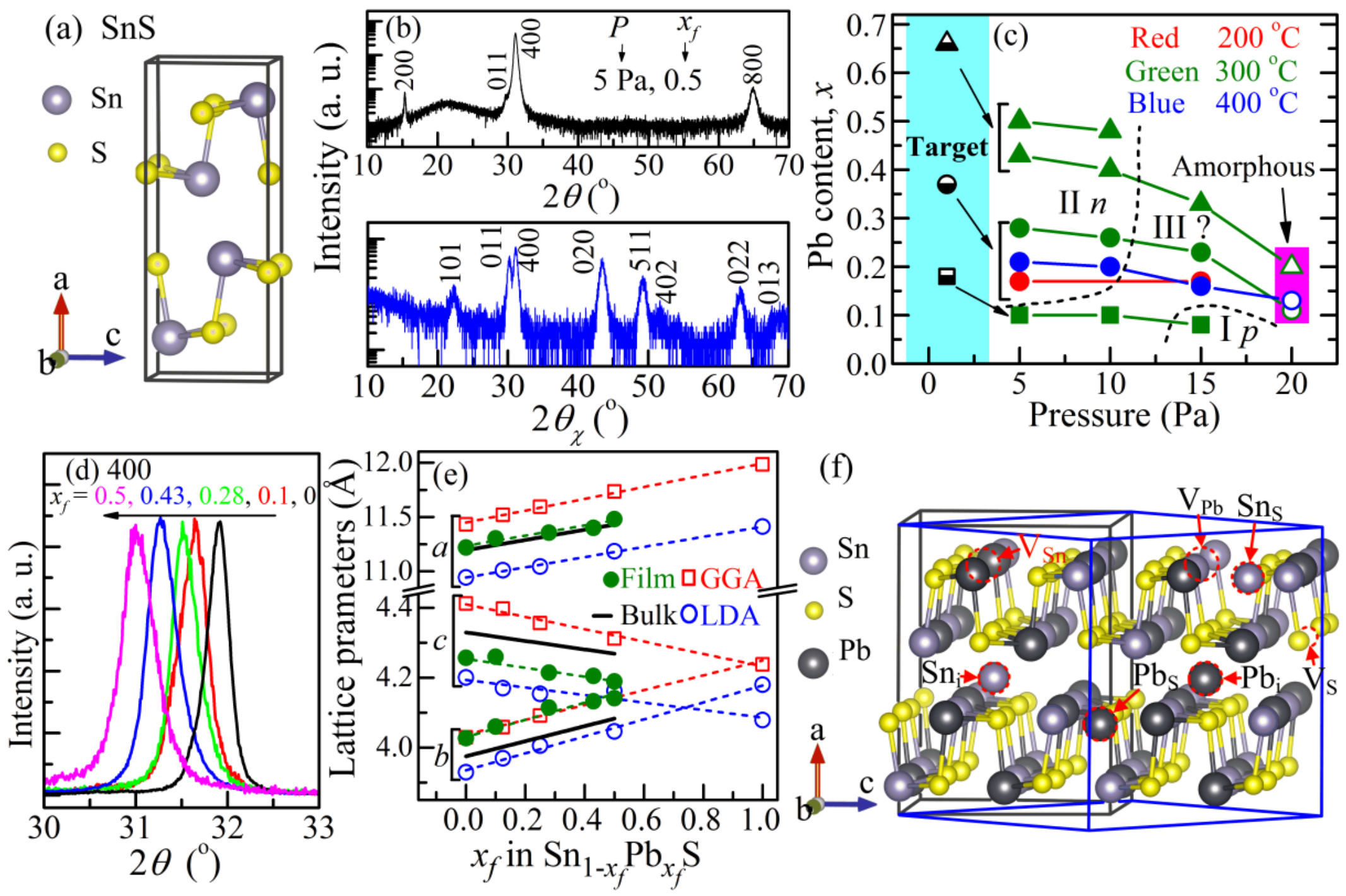

**Figure 1.** (a) Crystal structure of pure SnS. (b) Out-of-plane (top panel) and in-plane (bottom panel) synchronous scan XRD patterns of $(Sn_{1-x_f}Pb_{x_f})S$ film with $x_f$ = 0.5 grown at 300 °C and 5 Pa. (c) Pb content ($x_t$ for target, $x_f$ for thin films) as a function of pressure, substrate temperature, and $x_t$. The half-filled symbols indicate the targets, the closed circles the orthorhombic phase crystalline films, and the open symbols amorphous films. (d) 400 out-of-plane XRD diffraction peaks of the $(Sn_{1-x_f}Pb_{x_f})S$ films grown at 300 °C with various $x_f$ values. (e) Lattice parameters ($a$, $b$, $c$) of $(Sn_{1-x_f}Pb_{x_f})S$ films as a function of $x_f$. The closed circles indicate those obtained with the thin films, the solid lines are those of bulk $(Sn_{1-x}Pb_x)S$ taken from ref. 25, and the open symbols are the calculation results obtained by DFT in this work. The dashed straight lines are guides for eyes. (f) $(Sn_{1-x}Pb_x)S$ supercell model used for DFT calculations. The black line box draws the $(Sn_{16-n}Pb_n)S_{16}$ supercell model used for calculating the lattice parameters in **Figure 1**e. The blue line box draws the $(Sn_{32-n}Pb_n)S_{32}$ supercell model used for defect calculations in **Figure 3**, where the intrinsic defect models examined in this study are indicated also in the figure.

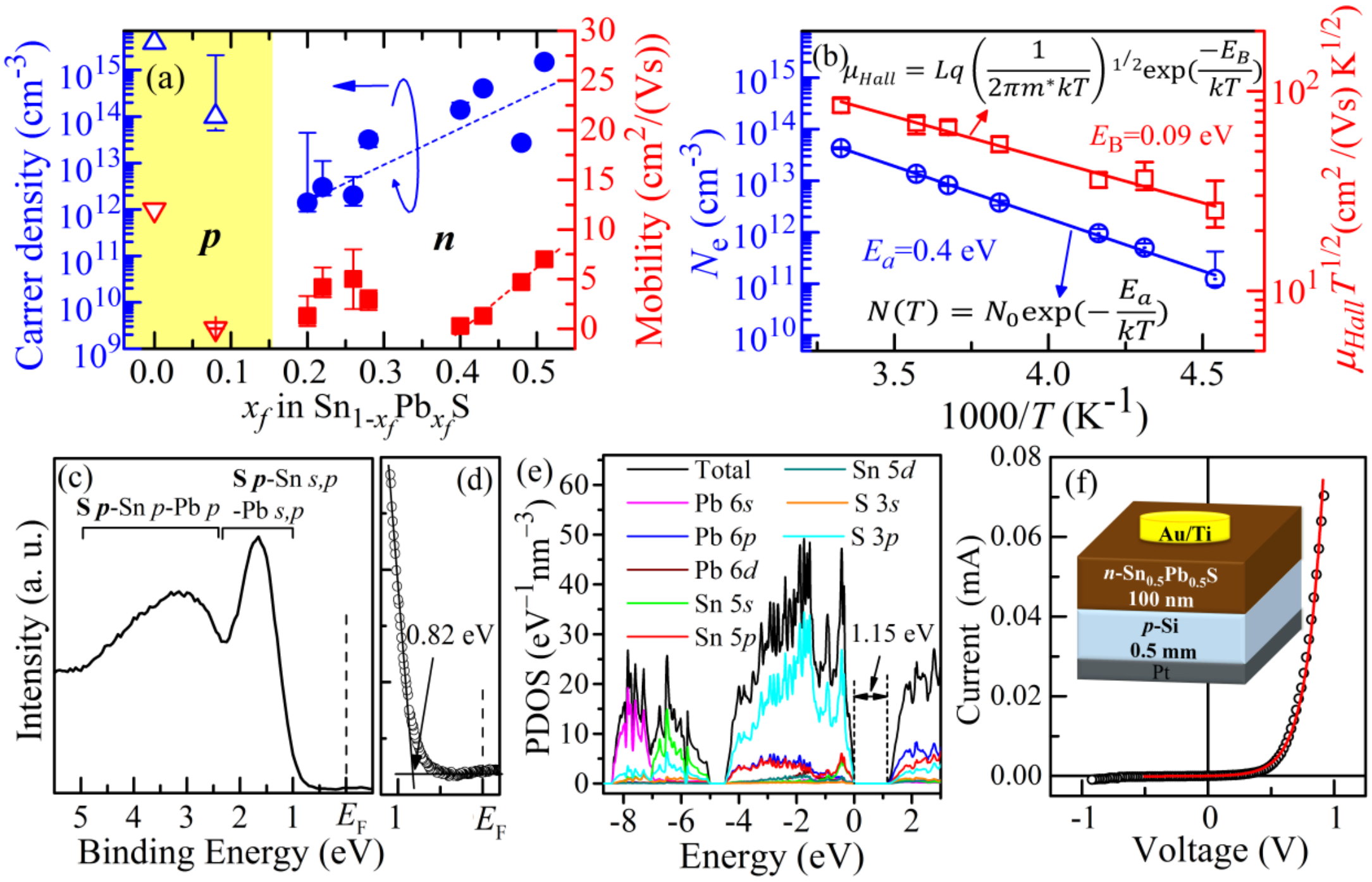

**Figure 2.** (a) Carrier density and mobility of ($Sn_{1-x_f}Pb_{x_f}$)S films as a function of $x_f$ measured by Hall effect. (b) Temperature dependences of electron density ($N_e$, blue line) and mobility ($\mu_e$, red line) of $n$-type ($Sn_{1-x_f}Pb_{x_f}$)S film with $x_f$ = 0.48. (c,d) UPS spectrum of $n$-type ($Sn_{0.5}Pb_{0.5}$)S film. (d) shows a magnified view near $E_F$. (e) Projected DOS of ($Sn_{0.5}Pb_{0.5}$)S calculated by DFT with GGA functionals. (f) I-V characteristics of $n$-($Sn_{0.5}Pb_{0.5}$)S/$p$-Si pn heterjunction. Inset shows the device structure.

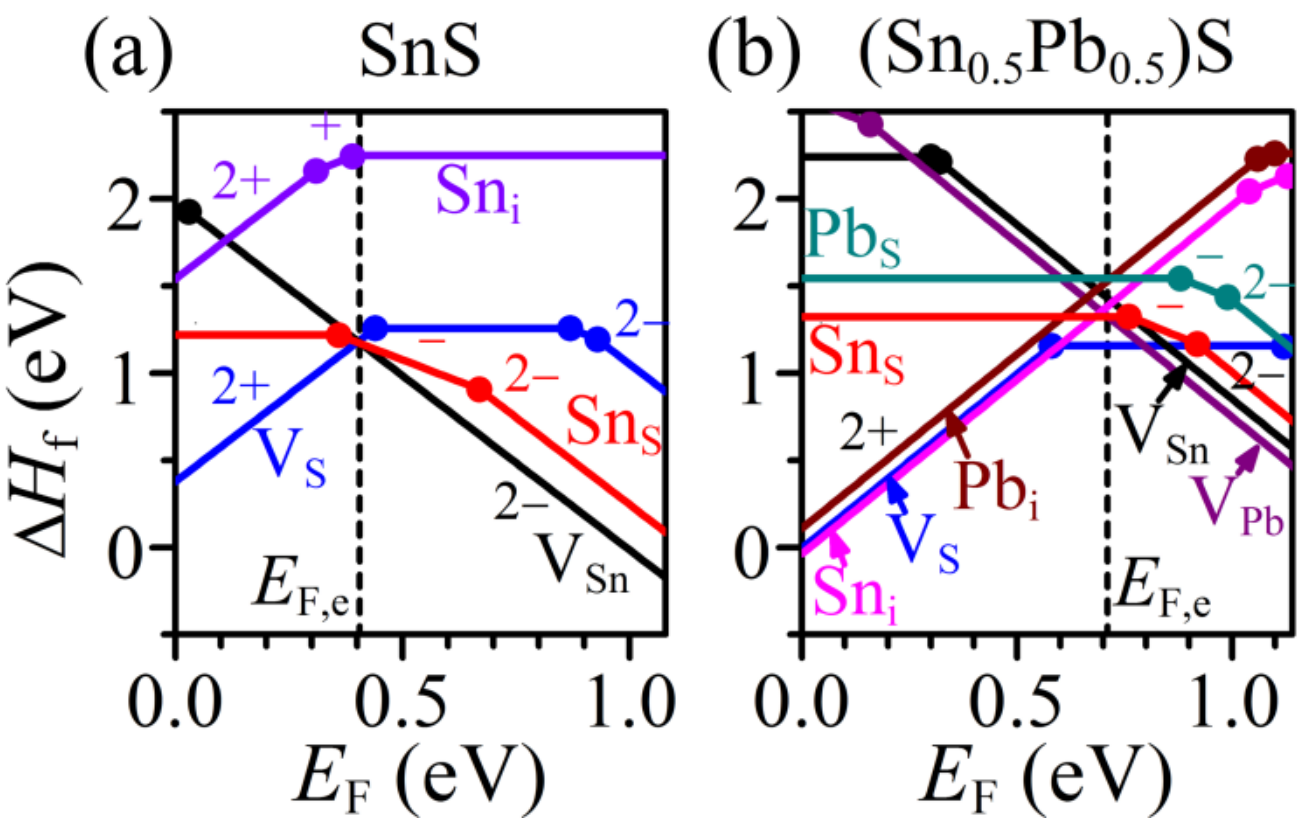

**Figure 3.** $\Delta H_f$ of intrinsic defects calculated at S-poor limit. Calculated $\Delta H_f$ for (a) pure SnS and (b) $(Sn_{16}Pb_{16})S_{32}$ models as a function of $E_F$ at S-poor limit. The values in the figures represent the charge states of the defects in the DFT calculations. The black dashed lines represent the equilibrium $E_F$ ($E_{F,e}$) calculated self-consistently at 300 °C.

**Carrier** polarity conversion from *p*-type to *n*-type was achieved for SnS thin films by isovalent ion substitution of $Pb^{2+}$ for $Sn^{2+}$, which is confirmed further by electrical characteristics of pn heterojunctions. The Pb substitution invokes a geometrical size effect and subsequently reduces the formation energies of Sn and Pb interstitials, which work as electron donors.

Fan-Yong Ran, Zewen Xiao, Yoshitake Toda, Hidenori Hiramatsu, Hideo Hosono, and Toshio Kamiya*

***N*-type Conversion of SnS by Isovalent Ion Substitution: Geometrical Doping As a New Doping Route**

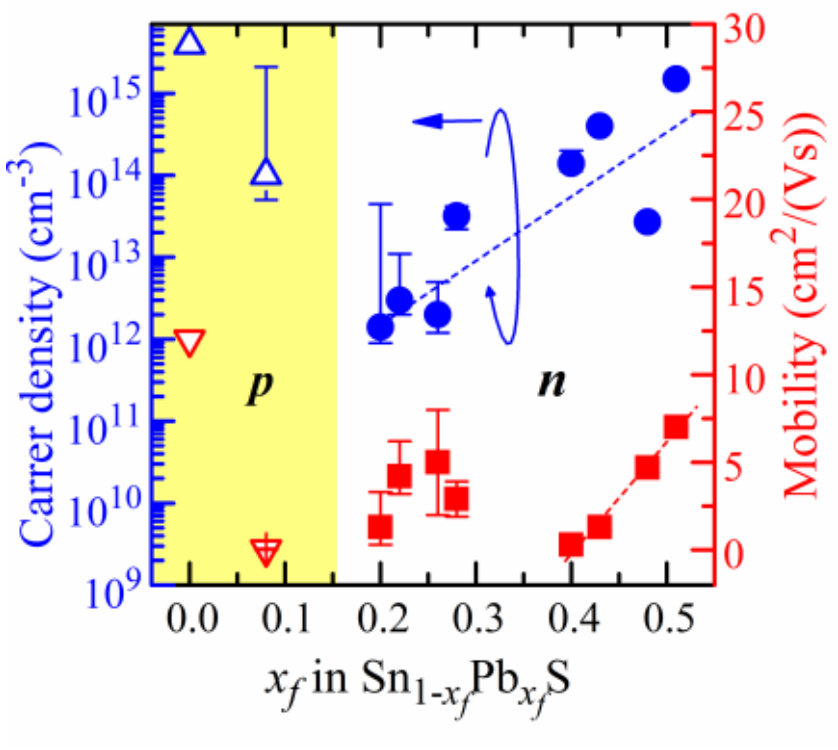

TOC figure

# Supporting Information

**_N_-type Conversion of SnS by Isovalent Ion Substitution: Geometrical Doping As a New Doping Route**

*Fan-Yong Ran, Zewen Xiao, Yoshitake Toda, Hidenori Hiramatsu, Hideo Hosono, and Toshio Kamiya**

# 1. Experimental Section

## 1.1. Film fabrication

$(Sn_{1-x_f}Pb_{x_f})S$ films of 100−200 nm in thickness were grown on $SiO_2$ glass substrate by pulsed laser deposition (PLD) using a KrF excimer laser (248 nm in wavelength, 3−6 J/cm$^2$ of laser energy density, and 10 Hz of repetition rate) with $(Sn_{1-x_t}Pb_{x_t})S$ polycrystalline targets in a $H_2S$ gas flow to control the S stoichiometry. The base pressure of the growth chamber was $1\times10^{-5}$ Pa. $T_s$ was varied from 200 to 400 $^oC$, and $P$ of an Ar / $H_2S$ mixing gas (80 / 20 %) from 5 to 20 Pa.

## 1.2. Characterization

The crystalline phase and crystal structure of the obtained films were characterized by X-ray diffraction (XRD, radiation source = Cu Kα). Optical properties were obtained by measuring transmittance ($T_r$) and reflectance ($R$) spectra. The absorption coefficient ($\alpha$) was estimated by $\alpha = \ln[(1-R)/T_r] / d$, where $d$ is the film thickness. Electrical properties of the SnS films were analyzed by Hall effect measurements using the van der Pauw configuration with an AC modulation of magnetic field. The Pb content in the films ($x_f$) were determined by X-ray fluorescence (XRF) spectroscopy calibrated by the chemical compositions obtained by inductively-coupled plasma-atomic emission spectroscopy (ICP-AES). The valence band structures were observed by UPS (excitation source = He I, 21.2 eV), where the films were protected in an Ar atmosphere during the transfer from the PLD chamber to the ultraviolet photoemission spectroscopy (UPS) chamber. The valence state of Pb was examined by X-ray photoemission spectroscopy (XPS, Mg Kα).

## 1.3. Calculation

Stable crystal / defect structures, their electronic structures, and formation energies of intrinsic defects were calculated by density functional theory (DFT) calculations with local density approximation (LDA) and generalized gradient approximation (GGA) functionals using the Vienna ab initio Simulation Package (VASP 5.3.3).[S1] The plane wave cutoff energy was set to 323.3 eV. A 32-atoms supercell model ($(Sn_{16-n}Pb_n)S_{16}$, black line in Figure 1f) and a 4×6×5 *k*-mesh were used for the calculations of structural properties and electronic structures. The defect calculations were performed using a 64-atoms model ($(Sn_{32-n}Pb_n)S_{32}$, blue line in Figure 1f) and a 3×3×3 *k*-mesh. The procedure for calculating the defect $\Delta H_f$ along with the general corrections followed the methodology reviewed by Zunger *et al*.[S2,S3] The equilibrium Fermi levels ($E_{F,e}$) were determined using the calculated density of states (DOS) by solving semiconductor statistic equations self-consistency so as to satisfy the charge neutrality condition.[S4]

## 2. XRD patterns of $(Sn_{1-x_f}Pb_{x_f})S$ films

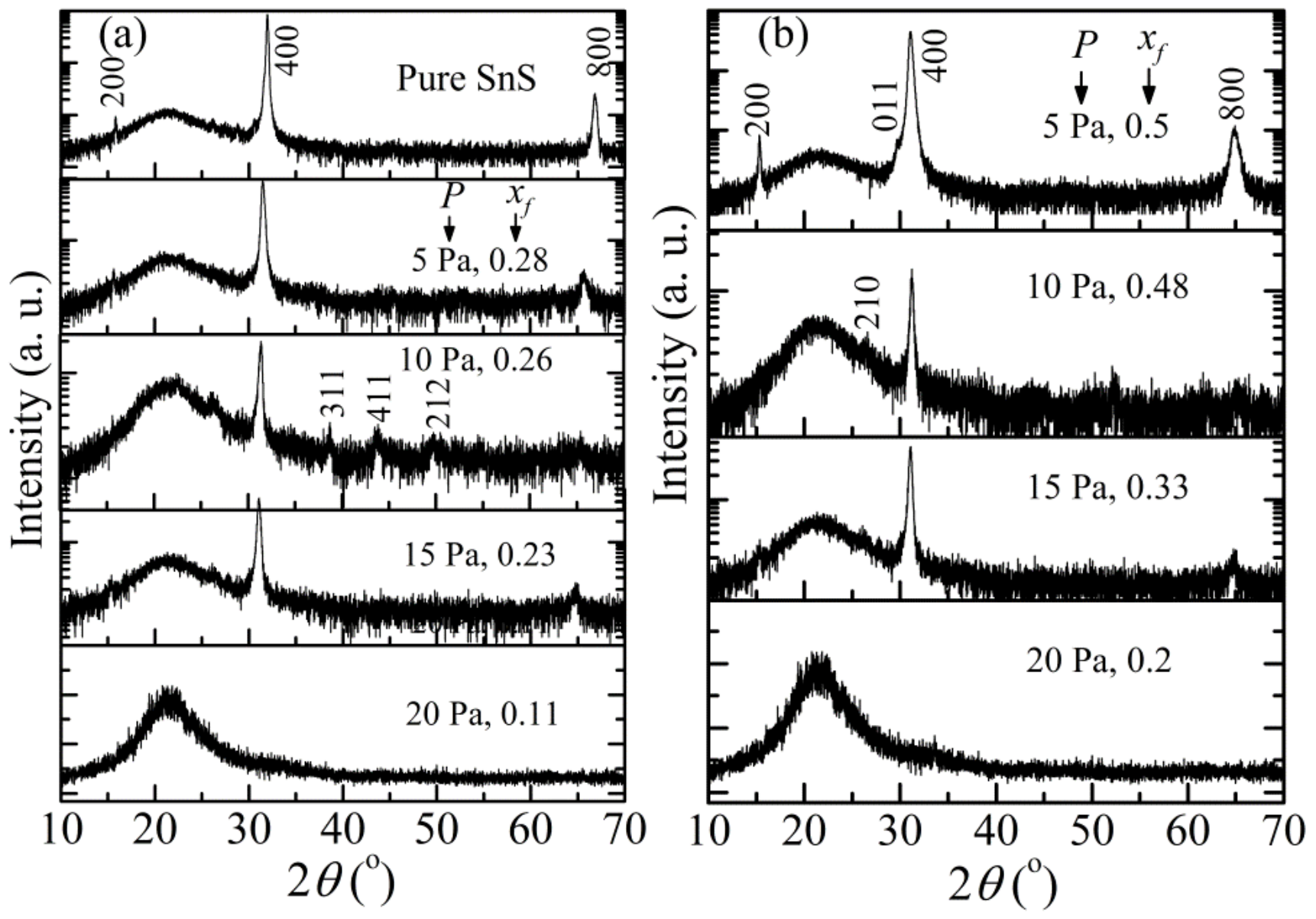

**Figure S1.** (a) and (b) XRD patterns of pure and $(Sn_{1-x_f}Pb_{x_f})S$ films fabricated at 300 °C at various conditions.

Typical out-of-plane XRD patterns of $(Sn_{1-x_f}Pb_{x_f})S$ films fabricated at various deposition conditions are shown in **Figure S1**. The XRD pattern of a pure SnS film is also shown at the top panel of **Figure S1**a for comparison. The XRD patterns of the films grown at $P \leq 15$ Pa exhibited strong 200, 400 and 800 diffractions from the orthorhombic structure of SnS. Whereas, no obvious diffraction was observed for the films grown at 20 Pa, indicating the films were amorphous.

## 3. Optical Properties

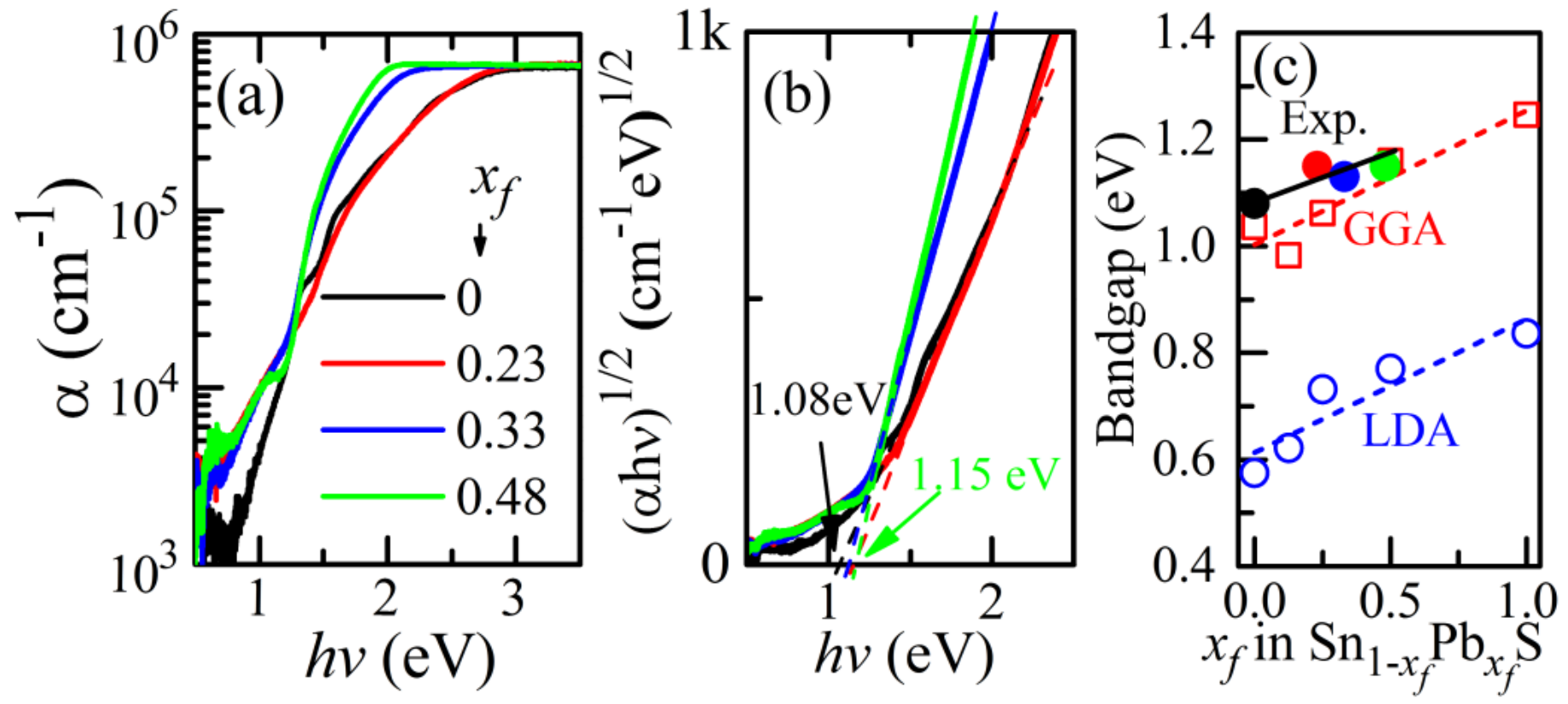

**Figure S2.** Band alignment of *n*-type $(Sn_{0.5}Pb_{0.5})S$ film with respect to other semiconductors.

**Figure S2**a and b show typical optical absorption spectra and $(\alpha h\nu)^{1/2} - h\nu$ plots (the indirect-transition model) of the $(Sn_{1-x_f}Pb_{x_f})S$ films fabricated at the various conditions, respectively. The pure SnS film (the black line in (a)) exhibited very weak subgap absorption, and the bandgap estimated from the $(\alpha h\nu)^{1/2}$–$h\nu$ plot is 1.08 eV, agreeing well with the literature theoretical value of ~1.07 eV.[S9] The bandgaps estimated from (b) are shown in **Figure S2**c as a function of $x_f$, showing that the bandgap value increased with $x_f$. Comparing with the calculated bandgap values, it is concluded that the GGA functionals reproduces the experimental values better than LDA.

## 4. Band Alignment of *n*-type $(Sn_{0.5}Pb_{0.5})S$ / *p*-type Si pn Junction

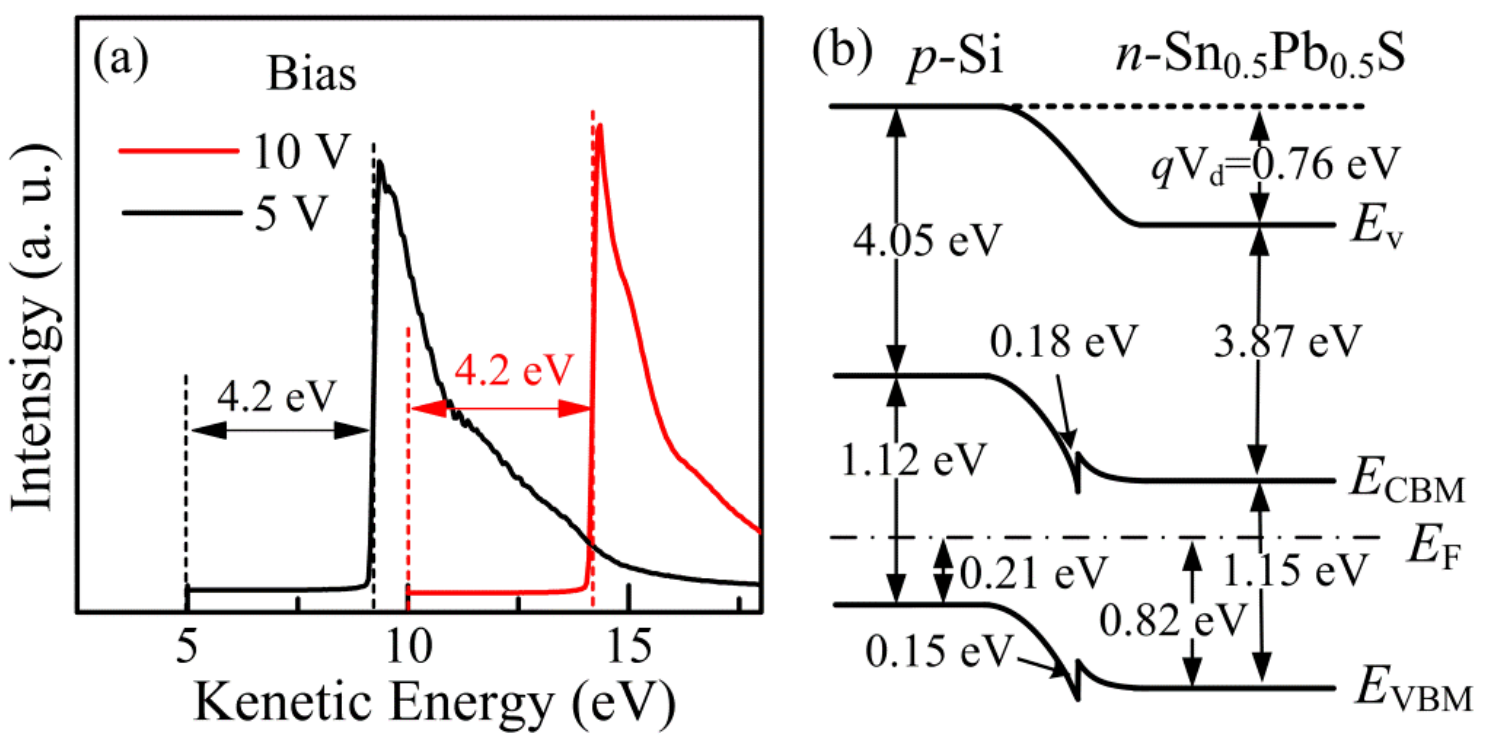

**Figure S3.** (a) UPS spectrum from $(Sn_{0.5}Pb_{0.5})S$ film with bias voltages of 5 and 10 V. (b) Band alignment of *n*-type $(Sn_{0.5}Pb_{0.5})S$ / *p*-type Si pn junction.

Band alignment of *n*-$(Sn_{0.5}Pb_{0.5})S$ / *p*-Si pn heterjunction is shown in **Figure S3**b, where the work function of the $(Sn_{0.5}Pb_{0.5})S$ film (4.2 eV) was determined from the relative cutoff energy of the secondary electrons in the UPS He I spectrum shown in **Figure S3**a and the electron affinity of Si (4.05 eV) obtained from a literature.[S5] It gives the built-in potential ($V_{bi}$) of 0.76 eV, the conduction band offset of +0.18 eV, and the valence band offset of +0.15 eV (the positive sign indicates that the band edges are higher in the $(Sn_{0.5}Pb_{0.5})S$ layer). The $V_{bi}$ roughly explains the experimental threshold voltage ~0.67 V (obtained by extrapolating the straight line region in **Figure 3**e).

## 5. Pb Valence State Examined by XPS

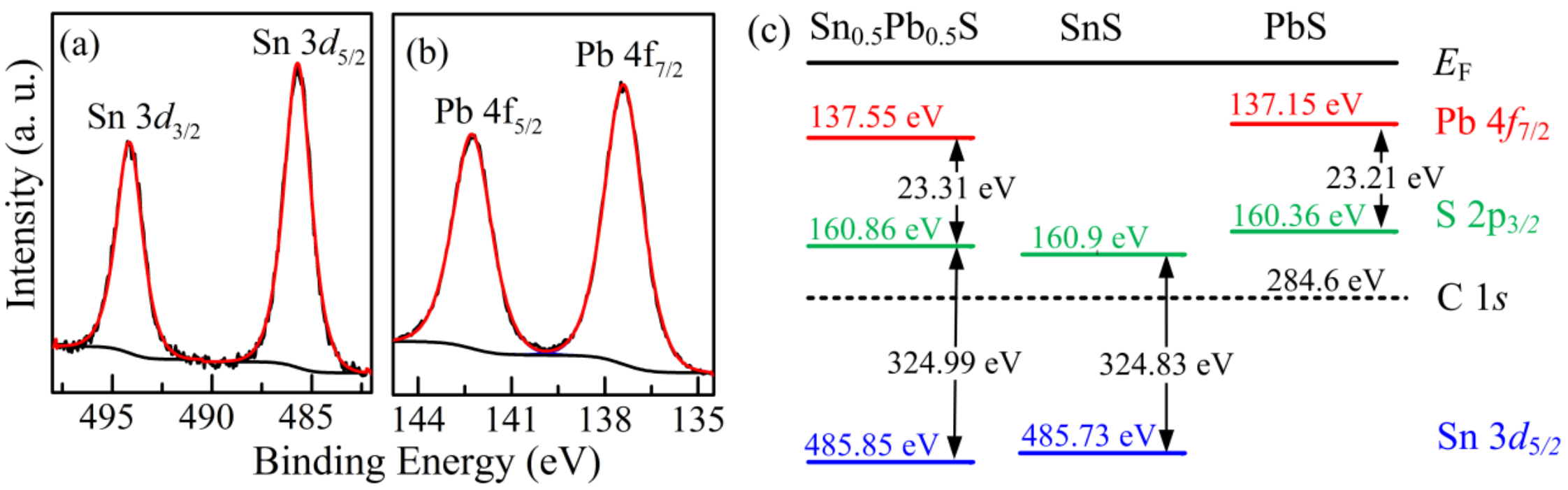

**Figure S4.** (a) Sn 3*d* and (b) Pb 4*f* core level XPS spectra of ($Sn_{0.5}Pb_{0.5}$)S film. (c) Energy alignment diagram for Sn 3*d*, Pb 4*f* and S 2*p* core levels of ($Sn_{0.5}Pb_{0.5}$)S film, pure SnS film, and PbS bulk sample.

XPS spectra were measured for ($Sn_{0.5}Pb_{0.5}$)S film, pure SnS film, and PbS bulk sample, where the samples were protected in an Ar atmosphere during the transfer from PLD preparation chamber/glove box to the XPS chamber. The binding energies were calibrated with reference to the C 1*s* level of adsorbed carbon-related molecules. Except for a little amount of OH, no other oxygen signal can be observed for all the samples. After cleaned by $Ar^+$ sputtering, the signal of OH disappeared, suggesting that the OH absorbed only on the surface. Sn 3*d* and Pb 4*f* core level spectra of the ($Sn_{0.5}Pb_{0.5}$)S film are shown in **Figures S**4a and b, respectively. The spectra were fitted using Gauss-Lorentz functions, and only one signal peak was obtained for each Sn 3*d* and the Pb 4*f* spectrum, indicating that the Sn and the Pb had single charge states, respectively.

The obtained energies of Sn 3*d*, Pb 4*f* and S 2p core levels of the ($Sn_{0.5}Pb_{0.5}$)S film, pure SnS film, and PbS bulk sample are aligned in the **Figure S**4c. The pure SnS film exhibited the Sn $3d_{5/2}$ peak at 485.73 eV, the energy difference between the Sn $3d_{5/2}$ peak and the S $2p_{3/2}$ peak ($\Delta_{Sn\text{-}S}$) was about 324.83 eV. The Pb $4f_{7/2}$ peak of the PbS sample was centered at 137.15 eV, the difference between the Pb $4f_{7/2}$ peak and the S $2p_{3/2}$ peak ($\Delta_{Pb\text{-}S}$) was 23.21 eV. These results are consistent with other reports on SnS and PbS samples.[S6-S8] For the ($Sn_{0.5}Pb_{0.5}$)S film, the Sn $3d_{5/2}$ peak was located at 485.85 eV with $\Delta_{Sn\text{-}S}$ of 324.99 eV, agreeing with those of the pure SnS. The Pb $4f_{7/2}$ peak was located at 137.55 eV, and $\Delta_{Pb\text{-}S}$ was 23.31 eV, very close to that of the PbS sample, suggesting that the Pb in the ($Sn_{0.5}Pb_{0.5}$)S film was of the +2 charge state.